\documentclass[superscriptaddress,prb,letter,aps,cp,amsmath,amssymb,reprint]{revtex4-2}

\usepackage{lmodern}
\usepackage{graphicx}
\usepackage{dcolumn}
\usepackage{bm}
\usepackage[utf8]{inputenc}
\usepackage[T1]{fontenc}
\usepackage{color}
\usepackage[colorlinks,citecolor=blue,linkcolor=blue,urlcolor=blue,bookmarks=false,hypertexnames=true]{hyperref}
\usepackage[inkscapelatex=false]{svg}

\begin{document}
	
	\newcommand{\fraunhoferIAF}{\affiliation{Fraunhofer Institute for Applied Solid State Physics IAF, Tullastr. 72, 79108 Freiburg, Germany}}
	\newcommand{\fraunhoferIWM}{\affiliation{Fraunhofer Institute for Mechanics of Materials IWM, Wöhlerstr. 11, 79108 Freiburg, Germany}}
	\newcommand{\inatech}{\affiliation{Department of Sustainable Systems Engineering (INATECH), University of Freiburg, Emmy-Noether-Str. 2, 79110 Freiburg, Germany}}
	
	\author{Philip Schätzle}\fraunhoferIAF\inatech
	\author{Reyhaneh Ghassemizadeh}\fraunhoferIWM
	\author{Daniel F. Urban}\fraunhoferIWM
	\author{Thomas Wellens}\fraunhoferIAF
	\author{Peter Knittel}\fraunhoferIAF
	\author{Florentin Reiter}\fraunhoferIAF
	\author{Jan Jeske}\fraunhoferIAF
	\author{Walter Hahn}\email{walter.hahn@iaf.fraunhofer.de}\fraunhoferIAF
	
	
	\title{Spin coherence in strongly coupled spin baths in quasi-two-dimensional layers}
	
	\begin{abstract} 
		We investigate the spin-coherence decay of NV$^-$-spins interacting with the strongly coupled bath of nitrogen defects in diamond layers. For thin diamond layers, we demonstrate that the spin-coherence times exceed those of bulk diamond, thus allowing to surpass the limit imposed by high defect concentrations in bulk. We show that the stretched-exponential parameter for the short-time spin-coherence decay is governed by the hyperfine interaction in the bath, thereby constraining random-noise models. We introduce a method based on the cluster-correlation expansion applied to strongly interacting bath partitions instead of individual spins. Our results facilitate material development for quantum-technology devices.
	\end{abstract}
	
	\maketitle
	
	Spin coherence is key for quantum-technology applications but the interaction of a spin with the surrounding bath typically leads to the loss of its spin coherence, i.e., its transverse polarization~\cite{slichter,abragam}. Generally, the dynamics in the bath creates a noisy magnetic field for the spin and dephases it. For weakly coupled spin baths, where the dynamics of the bath is slow compared to the central-spin dynamics, suitable theoretical approaches include (Gaussian) random-noise models~\cite{AndersonWeiss,KlauderAnderson,zhidomirov,de_lange_science,Witzel2014}, approximate analytical descriptions~\cite{das_sarma,PhysRevB.74.195301,hanson_coherent_2008,cole,Jeske_2012,Jeske_2013} and numerical simulations~\cite{PhysRevB.74.035322,PhysRevB.75.125314,hanson_coherent_2008,Yang2008,maze_decoh,Yang2017,Jeske_2013_2}. In strongly coupled spin baths, however, where the intra-bath and the center-to-bath interactions are of similar strength, the nonequilibrium many-body spin-coherence dynamics is not easily accessible by analytical or numerical techniques. The reasons are the absence of small parameters for a perturbative series expansion, the exponential scaling of resources for exact numerical simulations and the complex, even non-Markovian, bath dynamics. Such central-spin problems become increasingly important due to their relevance for spin-based quantum technologies~\cite{atature2018material,dowling2003quantum,awschalom2018quantum,davis_probing_2023} as, for example, quantum sensing~\cite{sensing_review}.
	
	\begin{figure} [b]
		\includegraphics[width=0.95\columnwidth]{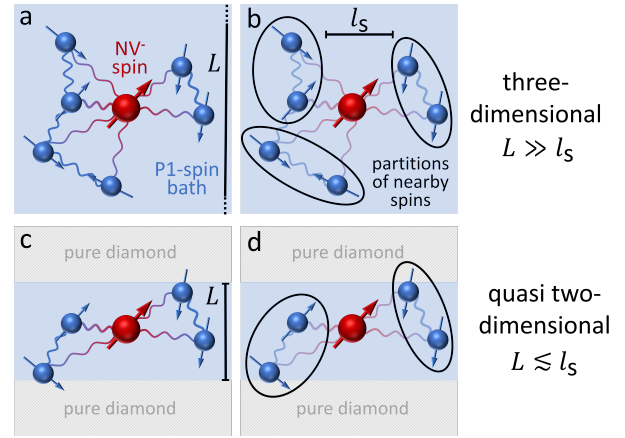}
		\caption{Partitioning the P1-center spin bath of different spatial dimensionality around the NV$^-$-center. Sketch of a three-dimensional (a) and quasi-two-dimensional (c) P1-center spin bath, which differ only by the layer thickness $L$. The relation between the layer thickness $L$ and the mean nearest-neighbor spin distance $l_\text{s}$ determines the dimensionality of the bath: $L\gg l_\text{s}$ for the three-dimensional and $L\lesssim l_\text{s}$ for the quasi-two-dimensional case. Wavy lines indicate the interaction between the spins. For the pCCE method, we first partition the bath into local strongly interacting indivisible partitions. As an example, two-spin partitions are indicated by the ovals in parts (b) and (d). Then, the conventional CCE method is applied to the partitions. Only the electron spin of the P1-centers is indicated in the figure, nuclear spins are omitted.}
		\label{fig_layers}
	\end{figure}
	
	For the nitrogen-vacancy (NV$^-$) center in diamond~\cite{DOHERTY2013}, which offers immense potential for quantum technologies~\cite{Wrachtrup_2006,review_meijer,one-second,27-spins,Yao2012,Morton2011,Maze2008,Barry2020,gaebel2006room,herbschleb2019ultra,taminiau2014universal,fault-tolerant,maile}, the spin-coherence decay is mainly governed by the interaction with spins of single-substitutional nitrogen defects (called $N_s^0$ or P1-centers)~\cite{hanson_coherent_2008} and the \textsuperscript{13}C nuclear spins~\cite{taminiau}. The P1-centers, each comprising an electron spin and a nuclear spin, usually dominate the spin-coherence decay of NV$^-$-centers even at comparatively low concentrations of about 1\,ppm due to the large electron-spin gyromagnetic ratio. To suppress the interaction with P1-centers, the material properties of diamond samples can be enhanced~\cite{Rodgers2021}, for example, by using delta-doping diamond-growth techniques~\cite{Ohno2012,Hughes2023,schatzle2023chemical} which allow for a precise depth placement of defects with layer thicknesses in the nanometer range. In such thin layers, the P1-center spin bath is typically quasi-two-dimensional, leading to longer Hahn-echo $T_2$ times~\cite{ye_spin_2019,Dwyer2D,davis_probing_2023,marcks2024}. The theoretical analysis of such quasi-two-dimensional strongly coupled baths is particularly difficult due to large fluctuations of dipolar interaction coefficients~\cite{feldman,lacelle,hahn_long-lived_2021}.
	
	In this letter, the results can be divided into three main parts: (I) In the effort to improve diamond samples for quantum-technology devices, we numerically simulate the spin-coherence decay of NV$^-$-spin ensembles in strongly coupled P1-center spin baths in diamond layers of different thickness and P1-center concentration, thereby identifying parameters for extended spin-coherence time. In particular, we calculate the spin-coherence times $T_2$ in quasi-two-dimensional diamond layers, when the mean nearest-neighbor spin distance $l_\text{s}$ satisfies $L\lesssim l_\text{s}$ (cf. Fig.~\ref{fig_layers}), and show that they exceed those in bulk diamond. (II) We show that the stretched-exponential parameter $p$ for the short-time spin-coherence decay is governed by the hyperfine interaction of the P1-centers, which strongly modifies the bath-spin dynamics (see, for example, Ref.~\cite{hanson_coherent_2008,park_decoherence_2022}). In particular, predictions from bath-dynamics models based on the Ornstein-Uhlenbeck process~\cite{davis_probing_2023} are valid only, when neglecting the nitrogen nuclear spins of the P1-centers in the numerical simulations. These results indicate the complexity of the spin dynamics in P1-center baths and offer direct means to identify more accurate effective magnetic-noise models for the description of P1-center spin baths~\cite{cappellaro_noise}. Moreover, these results are essential for efforts to suppress the magnetic noise generated by the spin bath.
	(III) To overcome difficulties of available methods when applied to strongly coupled baths of different dimensionality, we introduce a numerical method used in the above calculations, which we name partition-CCE (pCCE), cf. Fig.~\ref{fig_layers}. This method divides the spin bath into local strongly interacting subsystems, which we refer to as partitions and treat as indivisible units. Then, the cluster-correlation expansion (CCE)~\cite{Yang2008} is applied to the partitions instead of individual spins. In this way, the method takes into account important higher-order spin correlations leading to improved accuracy, broader range of applicability and enhanced convergence properties.
	
	
	
	We consider ensembles of NV$^-$-center electron spins (we omit the minus sign in the following) in isotopically purified $^{12}$C diamond layers with a sparse disordered P1-center spin bath of concentration $\rho_{\text{P1}}$. A strong magnetic field parallel to the NV-axis defines the $z$-axis. In the rotating reference frame, the P1-center electron spins interact via the secular part of the dipolar Hamiltonian~\cite{slichter,abragam}
	\begin{equation} \label{eqn_P1ham}
		{\cal H}_{\text{P1}}=\sum_{i>j\geq1}J_{ij}\left(I_{iz}I_{jz}-\frac{1}{4}[I_{i}^+I_{j}^-+I_{i}^-I_{j}^+]\right),  
	\end{equation}
	where $I_{jz}$ is the $z$-projection operator for the $j$-th P1-center electron spin, $I_{j}^+$ ($I_{j}^-$) is the corresponding spin raising (lowering) operator and \mbox{$J_{ij}\cong1/r_{ij}^3$} is the dipolar coupling constant with $r_{ij}$ being the distance between spins $i$ and $j$. For each P1-center, the hyperfine interaction with the nitrogen nuclear spin reads ${\cal H}_{\text{hf},j}=A_jI_{jz}S_{jz}$, where $S_{jz}$ is the spin-1 $z$-projection operator of the nuclear spin and $A_j$ is the hyperfine interaction constant, which depends on the Jahn-Teller-axis alignment~\cite{hanson_coherent_2008,park_decoherence_2022}. An energy mismatch generally suppresses the flip-flop transitions between the NV- and P1-center electron spins such that the corresponding interaction Hamiltonian reads~\cite{slichter,abragam} ${\cal H}_{\text{int}}=\sum_jJ_{0j}I_{0z}I_{jz}$, where the index 0 corresponds to the NV spin. We quantify the spin coherence of the NV spin by $\langle M_x(2\tau)\rangle\equiv\langle I_{0x}(2\tau)\rangle/\langle I_{0x}(0)\rangle$, where $\langle I_{0x}(2\tau)\rangle$ is the $x$-magnetization of the NV spin after a Hahn-echo sequence~\cite{hahn_echo}. For a more detailed description of the system, see supplemental material~\cite{suppl}.
	
	The interaction with the P1-center spin bath described by ${\cal H}_{\text{int}}$ leads to a time-dependent magnetic field for the NV spin $B^\text{P1}_z(t)\cong\sum_jJ_{0j}\langle I_{jz}(t)\rangle$, where $\langle I_{jz}(t)\rangle$ is the expectation value of $I_{jz}$. The characteristics of $B^\text{P1}_z(t)$ are governed by the dynamics of the P1-center bath, which is in turn dominated by the flip-flop transitions described in Eq.~\eqref{eqn_P1ham}. Generally, the Hahn-echo sequence cannot fully refocus an initial polarization of the NV spin for such time-dependent magnetic fields. Instead, the NV spin dephases in the $(x,y)$-plane corresponding to the decay of the spin coherence $\langle M_x(2\tau)\rangle$.
	
	The CCE method approximates the spin-coherence evolution $\langle M_x(2\tau)\rangle$ of a central spin in a large spin bath by including the contributions of \textit{all possible} spin clusters in the bath of size $n\leq N$ (CCE$N$) \mbox{$\langle M_x(2\tau)\rangle\approx\langle\langle\widetilde{\cal L}_{0}\widetilde{\cal L}_{1}\widetilde{\cal L}_{2}\cdots\widetilde{\cal L}_{N}\rangle\rangle,$} where $\langle\langle...\rangle\rangle$ denotes the average over mean-field spin configurations~\cite{suppl}, $\widetilde{\cal L}_{n}\equiv\prod_{i}\widetilde{\cal L}_{ni}$ ($i$ counts all possible $n$-spin clusters) and
	\begin{equation} \label{eqn_cce_gen}
		\widetilde{\cal L}_{ni}\equiv\frac{\langle M_x(2\tau)\rangle_{ni}}{\prod_{{\cal B}\subset{\cal C}_{ni}}\widetilde{\cal L}_{\cal B}},
	\end{equation}
	where ${\cal B}$ runs over all possible subclusters of ${\cal C}_{ni}$. $\langle M_x(2\tau)\rangle_{ni}$ is obtained by calculating the quantum dynamics only of spins in the cluster ${\cal C}_{ni}$, while all other bath spins are included on the mean-field level, i.e., they are static and point either up or down along the $z$-direction~\cite{Witzel2010,Witzel2012,onizhuk_probing_2021}.
	
	
	
	In practical calculations of the spin coherence of an NV electron spin in the weakly coupled bath of nuclear spins, CCE2 is often used (see, for example, Refs.~\cite{ye_spin_2019,seo_quantum_2016}), which includes the dynamics of single spins and spin pairs in the bath. However, in the strongly coupled P1-center spin bath, the dynamics in the bath takes place on the same timescale as the spin-coherence decay of the NV spin and, hence, correlated bath dynamics of higher order than for the nuclear-spin bath must usually be taken into account. Therefore, higher order CCE$N$ ($N>2$) need to be applied, which, however, often show unphysical behavior for strongly coupled spin baths~\cite{Witzel2010,Witzel2012}, when, for example, $|\langle  M_x(2\tau)\rangle|>1$ for an initially fully polarized NV spin. This can occur, for example, when clusters in the CCE method include \textit{only a part of a strongly interacting spin group}. In principle, this issue can be corrected for by further increasing $N$ but, in practice, this in turn introduces even more clusters with unphysical behavior.
	
	The main idea of the pCCE method is to divide the disordered spin bath into strongly interacting local subsystems (partitions) and to apply the conventional CCE method to the partitions instead of individual spins, cf. Fig.~\ref{fig_layers}. The partitions are chosen such that the interaction inside the partitions is maximized. These partitions act as indivisible units, when applying the CCE method~\cite{suppl}, and, thereby, facilitate capturing the correct spin dynamics. We choose for simplicity partitions with an equal number of spins $K$ (partition size) and adopt the notation pCCE($N,K$), when applying CCE$N$ after partitioning the bath into $K$-spin partitions. Using pCCE($N,K$), the clusters formed are unions of up to $N$ partitions. For example, in pCCE(2,2), CCE2 is applied to partitions of strongly interacting spin pairs as illustrated in Fig.~\ref{fig_layers}, and, in pCCE(2,4), each partition contains 4 spins such that the clusters considered are either the 4-spin clusters or the union of two 4-spin clusters~\cite{suppl}. Important to notice is also that pCCE($N$,1) is equivalent to CCE$N$.
	
	The pCCE method offers several advantages by construction. At sufficiently large $K$, each strongly-inter\-acting spin group is included in a single partition, thereby capturing important higher-order spin correlations. Moreover, in contrast to the CCE method, the number of samples for the mean-field average decreases with increasing $K$, cf. Fig.~\ref{fig_20tests} (for details about the mean-field average, see supplemental material~\cite{suppl}). Further, pCCE($N,K+1$) is not based on results of pCCE($N,K$) because the partitioning of the bath significantly changes. Therefore, to increase the accuracy of results at larger $K$, the pCCE method does not rely on correcting inaccuracies of results at smaller $K$. Hence, the order $K$ is effectively unrestricted. Another advantage is the significantly reduced number of clusters included. For example, in a simple comparison between pCCE($N,K$) and CCE$NK$~\cite{suppl}, the number of largest clusters of size $NK$ is $\binom{N_s/K}{N}$ and $\binom{N_s}{NK}$, respectively, where $N_s$ is the total number of bath spins. For example, for $N_s=80$, $N=2$ and $K=4$, the number of clusters is reduced by the factor $10^{-8}$.
	
	To demonstrate the performance of the pCCE method, $\langle  M_x(2\tau)\rangle$ for the NV spin interacting with a small disordered quasi-two-dimensional bath of 20 electron spins-1/2 is shown for different values of $K$ in Fig.~\ref{fig_20tests}. With increasing partition size $K$, the pCCE results converge towards the exact solution and become almost indistinguishable from the exact solution at $K=4$. This demonstrates the power of the pCCE method to capture the relevant dynamics and even approach the exact solution with a relatively low partition size $K$.
	
	Extensive tests of the pCCE($2,K$) performance for large spin baths show that $K=4$ is required for the method to be sufficiently accurate for all systems considered. The most demanding systems requiring $K=4$ are quasi-two-dimensional layers of P1-centers without the hyperfine interaction, i.e., electron spins-1/2, which implies that the correlated spin dynamics extends over local spin subsystems of at most $NK=8$ spins on the relevant timescales, thereby establishing an indirect measure of correlations in the bath. For P1-center spin baths (including the hyperfine interaction), pCCE converges already at $K=1$ corresponding to CCE2, which we attribute to the complex substructure in the bath suppressing flip-flop transition as discussed below (for convergence tests, see supplemental material~\cite{suppl}). We use $K=4$ for all systems in the following. In principle, the order $N$ could also be increased for pCCE but, for our calculations, this was not necessary. For the results described below, we perform disorder average over 100 randomly chosen spin distributions and average over 20 mean-field spin configurations for each system.

	\begin{figure}[t]
		\includegraphics[width=0.95\linewidth]{20_spin_example.png}
		\caption{Performance of the pCCE method. The time evolution of the spin coherence $\langle  M_x(2\tau)\rangle$ of the NV electron spin is shown for a quasi-two-dimensional bath of 20 randomly distributed electron spins-1/2. Lines connect points to guide the eye. When increasing the partition size $K$, pCCE($2,K$) converges towards the exact solution obtained using the Suzuki-Trotter expansion~\cite{suzuki} (for details, see supplemental material~\cite{suppl} and also Refs.~\cite{gemmer,canonical_typicality,popescu}). The number of samples for the mean-field average decreases for pCCE with larger $K$, whereas it increases for CCE with larger $N$. For example, the CCE3 results shown here have not converged after averaging over 1000 mean-field configurations, while pCCE(2,4) converged after averaging over 10 mean-field configurations. For details about the mean-field averaging, see supplemental material~\cite{suppl}.}
		\label{fig_20tests}
	\end{figure}
	
\begin{figure} [t]
		\includegraphics[width=0.99\columnwidth]{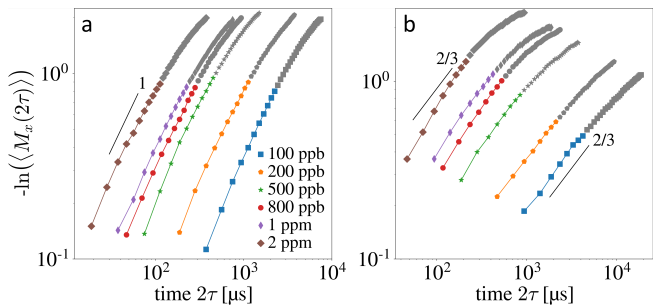}
		\caption{Evidence for stretched-exponential behavior of the spin coherence $\langle  M_x(2\tau)\rangle$ on short timescales. The function $-\ln(\langle  M_x(2\tau)\rangle)$ plotted on a double-logarithmic scale at layer thickness $L=240$\,nm (a) and $L=60$\,nm (b) for various P1-center concentrations $\rho_{\text{P1}}$ as indicated in the legend. The slope of the initial linear behavior corresponds to the stretched-exponential parameter $p$. In part (a), the layer thickness $L=240$\,nm corresponds to bulk diamond for all $\rho_{\text{P1}}$ considered ($p\approx1$). In part (b), the slope is changing corresponding to a transition from quasi-two-dimensional ($p\approx2/3$) towards three-dimensional bath when increasing $\rho_{\text{P1}}$ ($p>2/3$). The values used for fitting are indicated by the lines connecting the colored points. For $\langle  M_x(2\tau)\rangle$, this range approximately corresponds to the decay from 0.9 to 0.5. The standard deviation of the mean is smaller than the symbol size~\cite{suppl}.}
		\label{fig_loglog}
	\end{figure}

\begin{figure}[t!]
		\includegraphics[width=0.95\linewidth]{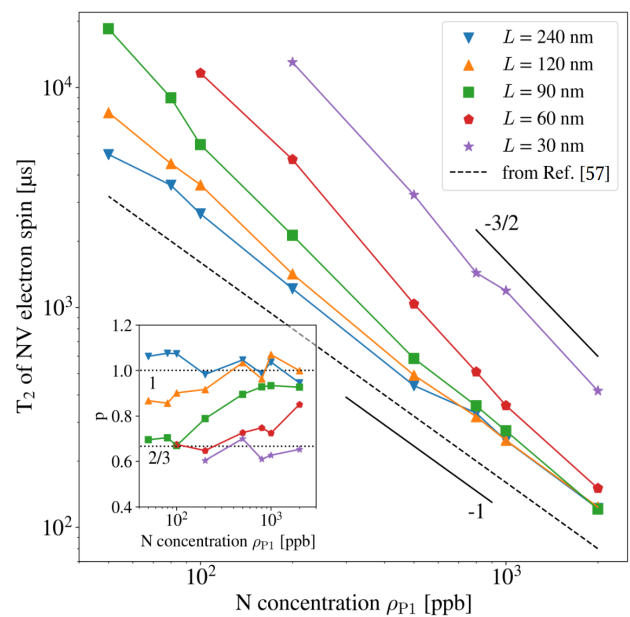}
		\caption{Coherence time $T_2$ and the stretch-exponential parameter $p$ signaling the spatial bath dimensionality. The $T_2$ time of the NV-center electron spin in a P1-center spin bath is shown for various P1-center concentrations $\rho_{\text{P1}}$ and layer thicknesses $L$ as indicated in the legend. Different scalings $T_2\cong(\rho_{\text{P1}})^{-1}$ and $T_2\cong(\rho_{\text{P1}})^{-3/2}$ are indicated by short black lines. The dashed black line is a fit to experimental results from Ref.~\cite{Bauch2020}. Inset: The corresponding stretched-exponential parameters $p$ with two distinguished values $p\approx1$ for bulk diamond and $p\approx2/3$ for quasi-two-dimensional layers. At intermediate $L$, we observe a smooth transition from the quasi-two-dimensional case ($p\approx2/3$) to the bulk diamond ($p\approx1$) when increasing $\rho_{\text{P1}}$. Lines connect points to guide the eye. Fluctuations are mainly due to uncertainties in the choice of the time window for fitting~\cite{suppl}.}
		\label{fig_T2_graph}
	\end{figure}

	Let us now discuss pCCE results for the spin-coherence $\langle  M_x(2\tau)\rangle$ decay of NV spins interacting with the P1-center spin bath. On short timescales~\cite{suppl}, $\langle  M_x(2\tau)\rangle$ is generally expected to follow a stretched-exponential function~\cite{KlauderAnderson} 
	\begin{equation} \label{eqn_stretched}
		\langle  M_x(2\tau)\rangle\approx\exp\left[-\left(\frac{2\tau}{T_2}\right)^p\right].
	\end{equation}
	For a well-defined stretched-exponential parameter $p$, the function $-\ln(\langle  M_x(2\tau)\rangle)$, when plotted on a double-logarithmic scale, is expected to exhibit a linear behavior with slope $p$. At short times, the results obtained indeed indicate a linear behavior, see Fig.~\ref{fig_loglog}. Fitting the function in Eq.~\eqref{eqn_stretched} to the numerical results in this regime, we obtain $p$ and $T_2$ for different values of the diamond-layer thickness $L$ and the P1-center concentration $\rho_{\text{P1}}$, see Fig.~\ref{fig_T2_graph}. There are two different scalings of the $T_2$ time as a function of $\rho_{\text{P1}}$, which we attribute to different spatial dimensionalities of the spin bath: $T_2\cong(\rho_{\text{P1}})^{-1}$ for bulk diamond at larger layer thicknesses (e.g. the slope is -1.02 for $L=240$\,nm) and $T_2\cong(\rho_{\text{P1}})^{-3/2}$ for quasi-two-dimensional layers (e.g. the slope is -1.46 for $L=30$\,nm)~\cite{hahn_long-lived_2021}. The quasi-two-dimensional nature of the $L=30$\,nm layer at the concentrations $\rho_{\text{P1}}$ considered is confirmed by relating $L$ to the mean nearest-neighbor spin distance $l_\text{s}=l_\text{s}(\rho_{\text{P1}})$: at $L=30$\,nm, $L/l_\text{s}$ varies approximately between 1.5 and 3 depending on $\rho_{\text{P1}}$ and, for $L=240$\,nm, $L/l_\text{s}>10$. For intermediate $L$, both regimes can be observed in Fig.~\ref{fig_T2_graph}: at larger $\rho_{\text{P1}}$, the spin bath tends to the three-dimensional case and, for smaller $\rho_{\text{P1}}$, to the two-dimensional case. These results highlight the broad range of applicability of the pCCE method and open new avenues to optimized material properties.

	\begin{table}[b]
		\caption{Evidence for the dependence of the stretched-exponential parameters $p$ on the hyperfine interaction in the P1-center bath. The parameter $p$ is calculated including (P1-center) and excluding (no hyperf.) the hyperfine interaction for various spin concentrations $\rho$. For the error estimate, see supplemental material~\cite{suppl}.}
		\begin{tabular}
			{|>{\centering}m{50pt}| >{\centering}m{45pt} >{\centering}m{45pt} >{\centering}m{45pt} >{\centering\arraybackslash}m{42pt}|}
			\hline
			&$\rho$&2\,ppm&1\,ppm&0.1\,ppm\\
			\hline
			
			two-dim.&P1-center&$0.65\pm 0.05$&$0.63\pm 0.05$&$0.67\pm 0.05$\\
			\cline{2-5}
			($L\leq$60\,nm)&no hyperf.&$0.95\pm 0.06$&$1.09\pm 0.07$&$1.05\pm 0.07$\\
			\hline
			three-dim. &P1-center&$0.95\pm 0.06$&$1.04\pm 0.06$&$1.08\pm 0.06$\\
			\cline{2-5}
			($L$=240\,nm)&no hyperf.&$1.62\pm 0.09$&$1.55\pm 0.09$&$1.51\pm 0.08$\\
			\hline
		\end{tabular}
		\label{table_e_P1}
	\end{table}
	
	The $T_2$ times obtained are consistent with experimentally measured values. We obtain for the coefficient $c$ in $T_2=c/\rho_{\text{P1}}$ for bulk diamond \mbox{($L=240$\,nm)} $c=253\pm13$\,µs\,ppm and the experimentally measured value is $160\pm12$\,µs\,ppm~\cite{Bauch2020, Barry2020}. The calculated $T_2$ times are generally considered upper bounds because of other potential sources of dephasing in experiments, such as other nitrogen-related defects~\cite{shinei2022nitrogen} or surface defects for shallow NV-centers~\cite{Dwyer2D}. Suppressing the coupling to such defects, for example, by means of decoupling pulse sequences or optimized sample growth, could lead to a further increase of the measured $T_2$ time. Further, for an accurate comparison between theory and experiment, the fitting procedures need also to be aligned~\cite{suppl}.
	
	
	The corresponding values for $p$ are shown in the inset of Fig.~\ref{fig_T2_graph}. We obtain \mbox{$p=1.03 \pm 0.03$} for bulk diamond ($L=240$\,nm) and $p=0.64 \pm 0.03$ for quasi-two-dimensional layers ($L=30$\,nm)~\footnote{Interestingly, the obtained values are consistent with analytical expectations for double electron-electron resonance (DEER) experiments~\cite{feldman,lacelle,hahn_long-lived_2021}.}. For intermediate diamond layer thicknesses $L$, we obtain a smooth transition between these two cases as we vary $\rho_{\text{P1}}$. Interestingly, if we leave out the nuclear spins of the P1-centers in our simulations, such that each P1-center is an electron spin-1/2, we obtain on average $p=1.56 \pm 0.04$ and $p=1.05 \pm 0.05$ for the bulk and the quasi-two-dimensional case, respectively, see Table~\ref{table_e_P1}. These values are consistent with those analytically predicted by bath-dynamics models based on the Ornstein-Uhlenbeck process~\cite{davis_probing_2023}. We attribute these findings to the complex internal substructure of the P1-center spin bath caused by the hyperfine interaction described by ${\cal H}_{\text{hf},j}=A_jI_{jz}{S}_{jz}$, where $A_j$ depends on the orientation of the Jahn-Teller axis of the P1-defects ($A_j=114$\,MHz for the [111]-axis and $A_j=86$\,MHz otherwise)~\cite{davies1981jahn}. The energy splittings induced by this interaction are much larger than the dipole-interaction constants at the concentrations $\rho_{\text{P1}}$ considered~\cite{suppl}. Therefore, the P1-center bath is effectively divided into five subgroups depending on the orientation of the nuclear spins and the Jahn-Teller axes. Flip-flop transitions between P1-center electron spins from different subgroups are strongly suppressed (see, for example, Ref.~\cite{hanson_coherent_2008,park_decoherence_2022}). Effectively, this situation corresponds to a NV spin in presence of five coupled spin baths with different characteristic times~\cite{suppl}. Even if the baths were uncoupled, the sum of Ornstein-Uhlenbeck processes describing the individual baths would typically not correspond to an Ornstein-Uhlenbeck process. We ascribe the dependence of the stretched-exponential parameter on the hyperfine interaction to this complexity in the bath.
	
	
	To conclude, we investigate the spin-coherence decay of NV$^-$-spins interacting with the strongly coupled disordered bath of the single-substitutional nitrogen defects in diamond layers. For quasi-two-dimensional diamond layers, we obtain $T_2$ times which exceed those in bulk diamond, hence allowing to surpass the limit imposed by the defect concentrations in bulk diamond. We show that the stretched-exponential parameter $p$ for the short-time spin-coherence decay is governed by the hyperfine interaction of the P1-centers, which is particularly important for understanding the underlying spin dynamics and for the effective modeling of the magnetic noise generated by the spin bath.
	We introduce the pCCE method based on the cluster-correlation expansion applied to local strongly interacting partitions of the bath, which includes important high-order spin correlations leading to better accuracy, applicability to longer timescales and enhanced convergence properties. We expect this method to be applicable beyond the setting considered here: for example, to NMR systems, i.e., nuclear spins on a lattice, to longer pulse sequences such as the Carr-Purcell-Meiboom-Gill pulse sequence and perhaps even to non-Markovian spin baths.
	
	The authors would like to thank V. V. Dobrovitski, G. A. Starkov and D. Maile for discussions of this work. PS and WH acknowledge funding from the executive-board project ``Quantum Computing'' - an initial project for the realization of a quantum computer of the Fraunhofer-Gesellschaft (QuaComVor). The source code is published in a GitHub repository~\cite{code}.

\newpage

\ 

\newpage
\onecolumngrid
\begin{center}
	{\bf SUPPLEMENTAL MATERIAL}
\end{center}

\setcounter{figure}{0}
\renewcommand{\thefigure}{S\arabic{figure}}

\setcounter{equation}{0}
\renewcommand{\theequation}{S\arabic{equation}}

\setcounter{table}{0}
\renewcommand{\thetable}{S\arabic{table}}

Note: In this supplemental material, we use the same notations as in the main article. Equation numbers, figure numbers and table numbers without an ``S'' refer to the main article.

\section{Detailed description of the system}
In this section, we discuss the NV-center spin system, the interaction between the NV center and the P1-center bath, describe details about the P1-center spin bath and, at the end, present the total Hamiltonian.

\subsection{NV center}
We consider ensembles of NV centers in [111]-oriented isotopically purified $^{12}$C diamond layers with a sparse disordered P1-center spin bath of concentration $\rho_{\text{P1}}$. Due to relatively long distances between the P1 centers, all defects can be treated as localized spins interacting via dipolar interaction. The diamond sample is placed in a strong magnetic field $B_z$ along the $z$-direction, which is also the [111]-direction, such that the three energy levels of the NV electron spin-1 are well separated (the NV-center nuclear spin is neglected, see below). It is important to note that the diamond layers are perpendicular to the [111]-direction and the magnetic field $B_z$. In presence of this magnetic field $B_z$, the internal Hamiltonian of the NV electron spin in the laboratory reference frame reads ${\cal H}^\text{lab}_{\text{NV}}=DS_z^2-\gamma B_zS_z$, where $D$ denotes the zero-field splitting, $S_z$ is the z-projection spin-1 operator and $\gamma$ the electron-spin gyromagnetic ratio. The NV spin is initialized in \mbox{$|\psi(0)\rangle=\frac{1}{\sqrt{2}}(|0\rangle+|-1\rangle)$, where $|0\rangle$ and $|-1\rangle$} are eigenstates of $S_z$ with corresponding eigenvalues. The state $|+1\rangle$ remains unoccupied on the timescales considered because the $z$-component of the NV spin is conserved by ${\cal H}^\text{lab}_{\text{NV}}$ and the interaction Hamiltonian with the P1-center spin bath $\cal H_\text{int}$ (see below). The NV spin can thus be effectively treated as a spin-1/2 with the adapted $z$-projection operator
\begin{equation}
	I_z=\begin{pmatrix}
		0 & 0\\
		0 & -1
	\end{pmatrix}.
\end{equation}
In the reference frame rotating with the Larmor frequency $\gamma B_z$, the NV-electron-spin Hamiltonian reads ${\cal H}_{\text{NV}}=-DI_{0z}$, where the index 0 refers to the NV electron spin.

We study the spin coherence of the NV spin, which we quantify by 
\begin{equation}
	\langle M_x(2\tau)\rangle\equiv\frac{\langle I_{0x}(2\tau)\rangle}{\langle I_{0x}(0)\rangle},
\end{equation}
where $\langle I_{0x}(2\tau)\rangle$ is the $x$-magnetization of the NV spin after a Hahn-echo sequence. A Hahn-echo sequence consists of the Hamiltonian time evolution of the system with time $\tau$, followed by a $\pi$-pulse along the $x$-axis for the NV spin and another Hamiltonian time evolution with time $\tau$~\cite{hahn_echo}. For the above initial state $|\psi(0)\rangle$, we obtain $\langle I_{0x}(0)\rangle=1/2$, such that $\langle M_x(2\tau)\rangle=2\langle I_{0x}(2\tau)\rangle$. Since the state $|\psi(0)\rangle$ corresponds to a fully polarized state in the Hilbert subspace described above, $\langle M_x(2\tau)\rangle\in[-1,1]$. Hence, we refer to any numerical result $|\langle M_x(2\tau)\rangle|>1$ as unphysical.

The $^{14}$N nuclear spin of the NV center can be assumed static on the timescales considered. The dominating interaction term in the strong magnetic field $B_z$ is of the form $I_z\tilde{S}_z$~\cite{slichter}, where $\tilde{S}_z$ is the $z$-projection operator of the nuclear spin. Such interaction would create an effective static magnetic field along the $z$-direction for the NV electron spin. Therefore, the effect of the NV nuclear spin on the NV electron spin is canceled in the Hahn-echo sequence. Hence, we do not include the $^{14}$N nuclear spin in our numerical simulations.

\subsection{Interaction between the NV spin and the P1-center spin bath}
The central spin and the P1-center bath spins interact by dipolar coupling with coupling constants \mbox{$J_{ij}=\mu_0\gamma^2(1-3\cos^2(\theta_{ij}))/4\pi|\vec{r}_{ij}|^3$}, where $\theta_{ij}$ is the angle between the vector $\vec{r}_{ij}$ connecting the spins and the axis of the applied magnetic field, which defines the $z$-axis. The magnetic field is usually chosen such that the NV electron spin and the P1-center electron spins are not in resonance. Therefore, the flip-flop terms in the dipolar Hamiltonian are suppressed and the only secular part remaining in the rotating reference frame is 
\begin{equation}
	{\cal H}_{\text{int}}=\sum_{j\geq1}J_{0j}I_{0z}I_{jz},
\end{equation}
where the index 0 corresponds to the NV electron spin and $j\geq1$ to P1-center electron spins.

The P1-center spins are assumed to be in the infinite-temperature limit, when each P1-center spin is described by a fully mixed density matrix 
\begin{equation}
	\rho=\begin{pmatrix}
		1/2 & 0\\
		0 & 1/2
	\end{pmatrix}.
\end{equation}

\subsection{\label{P1_bath}P1-center spin bath}

A P1 center consists of an electron spin-1/2 and a $^{14}$N nuclear spin-1. Between the P1 centers, the interaction is dominated by the dipolar interaction between the respective electron spins. The corresponding Hamiltonian ${\cal H}_\text{P1}$ is given in Eq.~(1) of the main text.

For each P1 center, the strong hyperfine interaction in the rotating reference frame is of the form
\begin{equation}
	{\cal H}_{\text{hf},j}=A_jI_{jz}S_{jz},
\end{equation}
where $A_j$ is the hyperfine interaction constant and $S_{jz}$ is the z-projection operator for the nuclear spin-1 of $j$-th P1 center~\cite{hanson_coherent_2008,park_decoherence_2022}. The hyperfine interaction constant $A_j$ depends on the orientation of the Jahn-Teller (JT) axis of the P1 center: for the [111]-axis (corresponding to the $z$-axis here), $A_j=114$\,MHz, and $A_j=86$\,MHz for the other three possible axis alignments. For the P1-center concentrations $\rho_\text{P1}$ considered here, the energy splittings induced by the hyperfine interaction are much larger than the dipole-interaction constants (see below) and, therefore, two P1-center electron spins can flip-flop only if the orientation of the corresponding nuclear spins and the JT axes is the same for the two P1 centers. Hence, the P1-center bath is effectively divided into five spin subgroups depending on the orientation of both, the JT axes and the nitrogen nuclear spins. Any P1 center remains in the same subgroup on the timescales of our simulations, because the nuclear spin dynamics as well as the JT reorientation have significantly longer characteristic timescales~\cite{davies1981jahn}. This complex structure of the P1-center bath implies that the majority of nearest neighbors of any P1-center spin is likely to belong to a different subgroup. These nearest neighbors create strong disordered local fields for the P1-center spin by means of the first Hamiltonian term in Eq.~(1) but do not flip-flop with it. 

The above P1-center bath subgroups are not equally sized. In order to understand this, let us first consider the individual configurations for the nuclear-spin polarization and the JT-axis alignment. The different $m_s$ states for the nuclear spins are $|+1\rangle$, $|-1\rangle$ and $|0\rangle$. For the first two states, there are the two inequivalent JT axes alignments: JT along [111] and JT not along [111]; we denote the latter case by $\overline{[111]}$ (it comprises the three equivalent alignments [$\overline{1}$11], [1$\overline{1}$1] and [11$\overline{1}$]). For the state $|0\rangle$, the hyperfine interaction term vanishes, thus the JT-axis alignment remains unimportant. For an even distribution among the nuclear-spin states (infinite temperature) and the JT alignments~\cite{davies1981jahn}, the probability $P$ for the different configurations is as follows: for $|+1\rangle$ and [111]: P=1/12, for $|-1\rangle$ and [111]: P=1/12, for $|+1\rangle$ and $\overline{[111]}$: P=3/12, for $|-1\rangle$ and $\overline{[111]}$ P=3/12, for $|0\rangle$: P=4/12 (in this case, the JT-axis alignment is unimportant). It is also noteworthy that the different sizes of the P1-center bath subgroups lead to different effective spin concentrations in the individual subgroups and, hence, to different characteristic times for the dynamics in these subgroups.

\begin{figure}[t]
	\includegraphics[width=0.6\linewidth]{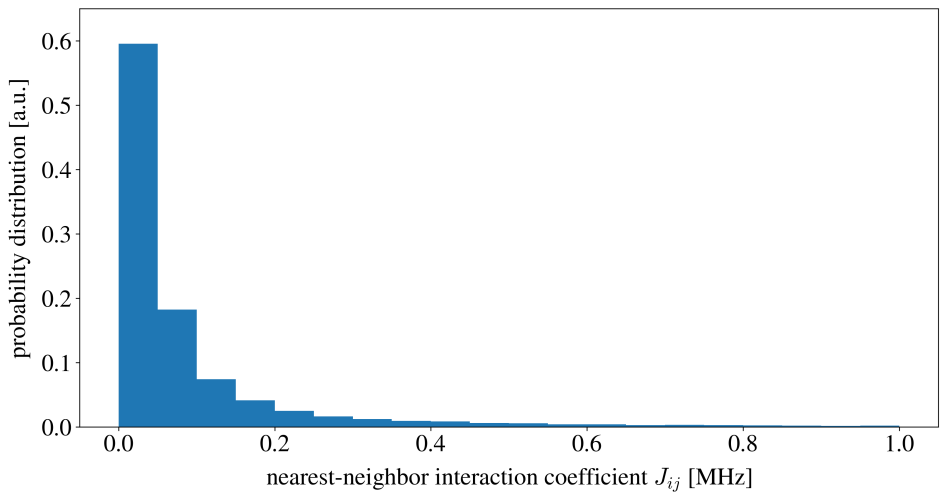}
	\caption{Histogram of nearest-neighbor interaction constants $J_{ij}$ in a three-dimensional diamond layer at the highest concentration $\rho_{\text{P1}}=2$\,ppm considered in this letter. The overwhelming majority of nearest-neighbor interactions constants is well below the smallest hyperfine energy splitting of 28\,MHz induced by different JT-orientations. At lower concentrations $\rho_{\text{P1}}$, the interaction constants $J_{ij}$ become even smaller. These results confirm the assumption, that the P1-center bath is effectively split into subgroups. For details, see the text.}
	\label{fig_app_JT}
\end{figure}

The validity of the assumption that spins from different subgroups do not flip-flop with each other is dependent on the P1-center concentration $\rho_{\text{P1}}$ because the dipolar interaction coefficients increase with larger concentration $\rho_{\text{P1}}$. The highest P1-defect concentration we consider in this work is $\rho_{\text{P1}}=2$\,ppm. We calculate the nearest-neighbor interaction constants for each spin in 100 randomly chosen systems at this concentration $\rho_{\text{P1}}=2$\,ppm; the resulting histogram is shown in Fig.~\ref{fig_app_JT}. The overwhelming majority of nearest-neighbor interactions is well below the smallest hyperfine energy splitting of 28\,MHz induced by different JT-orientations. At lower concentrations $\rho_{\text{P1}}$, the interaction constants $J_{ij}$ shift to even lower values. This corroborates our assumption, that spins from different subgroups do not flip-flop with each other.

\subsection{Total Hamiltonian}
Let us now put together all terms discussed above. The total Hamiltonian ${\cal H}_\text{tot}$ considered reads
\begin{equation}
	{\cal H}_\text{tot}={\cal H}_{\text{NV}}+\cal H_\text{int} +\cal H_\text{P1} +{\cal H}_\text{hf},
\end{equation}
where ${\cal H}_{\text{NV}}$ describes the NV center, $\cal H_\text{int}$ describes the interaction between the NV center and the P1-center bath, $\cal H_\text{P1}$ describes the interaction between the P1 centers and ${\cal H}_\text{hf}$ describes the hyperfine interaction for each P1 center. Substituting the individual terms in the expression above, the total Hamiltonian ${\cal H}_\text{tot}$ in the rotating reference frame reads
\begin{equation} \label{eqn_ham_tot}
	{\cal H}_\text{tot} =-DI_{0z}+\sum_{j\geq1}J_{0j}I_{0z}I_{jz}+\sum_{i>j\geq1}J_{ij}\left(I_{iz}I_{jz}-\frac{1}{4}[I_{i}^+I_{j}^-+I_{i}^-I_{j}^+]\right)+ \sum_{j\geq1}A_jI_{jz}S_{jz}.
\end{equation}
For the dynamics in a Hahn-echo sequence, the effect of the first term ${\cal H}_{\text{NV}}$ is canceled and, hence, this term can be left out.

The hyperfine interaction ${\cal H}_\text{hf}$ (last term in Eq.~\eqref{eqn_ham_tot}) describes effectively an additional constant magnetic field along the $z$-axis, since all $S_{jz}$ are conserved on the timescales considered here. These terms can be omitted by considering a rotating reference frame with rotation frequency dependent on the P1-center bath subgroups discussed above. The flip-flop Hamiltonian terms for two P1-center electron spins from different subgroups acquire a fast oscillating term in this rotating reference frame and, hence, these flip-flop terms are suppressed and can be left out. In our numerical simulations, we consider this rotating reference frame.

\section{Mean-field averaging}
As shown in the main article, the expansion for CCE$N$ reads
\begin{equation} \label{eqn_ccefinal}
	\langle M_x(2\tau)\rangle\approx\langle\langle\widetilde{\cal L}_{0}\widetilde{\cal L}_{1}\widetilde{\cal L}_{2}\cdots\widetilde{\cal L}_{N}\rangle\rangle,
\end{equation}
where $\langle\langle...\rangle\rangle$ denotes the average over mean-field spin configurations. For this average, a mean-field configuration is randomly chosen for each P1-center electron spin, i.e., either up or down along the $z$-direction. In the calculations of $\langle M_x(2\tau)\rangle_{ni}$, these mean-field spin configurations are chosen for all spins outside the cluster ${\cal C}_{ni}$ (for the definition of $\langle M_x(2\tau)\rangle_{ni}$ and ${\cal C}_{ni}$, see main article). The average over mean-field configurations is performed for the total factor on the right-hand-side in Eq.~\eqref{eqn_ccefinal}. We refer to this mean-field average as the normal average in the following.

Another type of the mean-field bath-spin averaging is often used for the suppression of unphysical CCE results (see, for example, Refs.~\cite{Witzel2010,Witzel2012,onizhuk_probing_2021}), which we refer to as internal mean-field averaging in the following. Internal mean-field averaging is defined as follows: each term $\widetilde{\cal L}_{ni}$ (for definition, see main article) is averaged over mean-field spin configurations separately before the individual factors are put together in Eq.~\eqref{eqn_ccefinal}. When using the internal averaging, correlations between the individual factors on the right-hand-side of Eq.~\eqref{eqn_ccefinal} are neglected. However, for disordered spin baths, these correlations seems to be negligible in practical calculations, cf. Fig.~(2).

The internal averaging is particularly useful in CCE calculations with many unphysical terms, i.e., many clusters with $\langle M_x(2\tau)\rangle_{ni}>1$. When using internal averaging, the numerator and each term in the denominator in Eq.~(2) are averaged separately and this leads to the suppression of unphysical terms. The two kinds of averaging (normal and internal) can be combined in the following way. When obtaining the result for $\langle M_x(2\tau)\rangle$ in Eq.~\eqref{eqn_ccefinal} with internal averaging, this calculation can be repeated many times, thereby averaging over results for $\langle M_x(2\tau)\rangle$. We use both types of mean-field averaging in this letter as explained below.

For Fig.~2 of the main text, the number of samples for normal and internal averaging is as follows: pCCE(2,1) 50 normal without internal, pCCE(2,2) 50 normal with 50 internal, pCCE(2,3) 50 normal with 20 internal, pCCE(2,4) 10 normal without internal, CCE3 1000 internal. For the calculation of the $T_2$ time and the stretched-exponential parameter $p$ in P1-center baths (with and without the hyperfine interaction), we use pCCE(2,4) with 20 samples for internal averaging. In general, for larger partition size $K$, the number of samples for the mean-field averaging decreases for pCCE, while it increases for CCE for larger $N$. This illustrates one of the advantages of the pCCE method. The reason for the decreasing number of samples for the mean-field averaging for the pCCE method is that the average over spins in the clusters is already included by the quantum-mechanical average and that the clusters consist of local partitions.

When increasing the partition size beyond $K=4$, we observed that, for $K=5$ and $K=6$, the number of samples for the mean-field average further decreases, so we expect that, for sufficiently large $K$, neither internal nor normal mean-field average is required. Using exact methods for the time propagation, pCCE(2,10) or even larger partition sizes are feasible.

\subsection{Internal mean-field averaging in the conventional CCE method}
\begin{figure}[b]
	\includegraphics[width=0.6\linewidth]{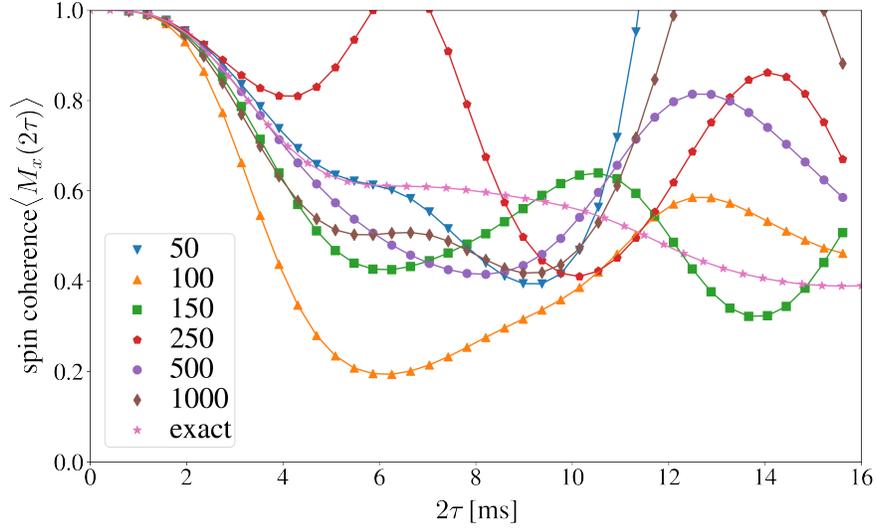}
	\caption{Results of CCE3 applied to the 20 electron spin-1/2 bath in a quasi-two-dimensional layer (from Fig.~(2)) for different numbers of samples for the internal average as indicated in the legend. At 1000 samples for the internal average, the CCE3 results have not converged and no signatures of convergence can be observed.}
	\label{fig_CCE3_small}
\end{figure}

We apply CCE3 to the bath of 20 electron spins-1/2 in a quasi-two-dimensional layer considered in Fig.~2 of the main article. We vary the number of samples for internal averaging to mitigate the unphysical behavior $\langle M_x(2\tau)\rangle>1$; the results are shown in Fig.~\ref{fig_CCE3_small}. The results imply that, with increasing number of samples for the internal mean-field average, no enhancement of the accuracy is achieved because the results do not converge even after averaging over 1000 samples. We conclude that this small bath of pure electron spins-1/2 in quasi two dimensions represents a challenging system for the conventional CCE method requiring an even larger number of samples for the mean-field average.

\begin{table}[t]
	\begin{tabular}{c | c}
		Number of spins&Occurrence of $\langle M_x(2\tau)\rangle$ > 1\\
		\hline
		5&0 \%\\
		10&28 \%\\
		15&26 \%\\
		20&40 \%\\
	\end{tabular}
	\caption{Statistics for the occurrence of unphysical behavior ($\langle M_x(2\tau)\rangle>1$) for CCE3. We consider a series of smaller systems that we obtain by retaining only the 5, 10 or 15 closest spins to the NV center from the system of 20 electron spins in a quasi-two-dimensional layer, shown in Fig.~2 of the main article. These calculations were repeated 50 times and the fraction indicated in this table refers to this number of repetitions.}
	\label{table_cce3}
\end{table}

We assume that the problem of unphysical behavior ($\langle M_x(2\tau)\rangle>1$) becomes worse with increasing system size because, in larger spin baths, even more clusters are formed, each of which can lead to $\langle M_x(2\tau)\rangle>1$. To challenge this assumption, we consider a series of smaller systems that we obtain by retaining only the 5, 10 or 15 closest spins to the NV center from the system of 20 electron spins considered in Fig.~\ref{fig_CCE3_small}. We use CCE3, fix the number of samples for the internal mean-field average to 250 and count the occurrence of unphysical behavior, i.e., $\langle M_x(2\tau)\rangle>1$, while repeating the same calculation with random choices of the mean-field bath-spin configurations. The probability for the occurrence of such behavior increases with larger system size as shown in Table~\ref{table_cce3}. 

The above results imply that, even if the conventional CCE method of high order $N$ yields accurate results for the system considered in this article, this would involve a very large number of samples for the mean-field average. The pCCE method allows to increase the order $K$ while reducing the number of samples for the mean-field average. If we were to use the conventional CCE approach such that it includes all terms entering the pCCE(2,4) order, this would require CCE8 leading to a huge number of samples for the internal mean-field average provided such high-order CCE converges, cf. Ref.~\cite{Witzel2012}.

\section{Partitioning method}
A key component of the proposed pCCE method is the partitioning of the spin bath into local strongly interacting subsystems (partitions). We choose for simplicity the bath partitions to be of the same size $K$, corresponding to the same number of spins in the partitions. The partitioning is performed using a variant of the k-means algorithm. However, the method in general also allows a flexible partitioning with partitions of different size.

Let us first discuss the conventional k-means algorithm. According to the k-means algorithm, first, $m$ central points at variable positions $\vec{y}_i$ are introduced and, second, the spins at positions $\vec{x}_j$ are assigned to the different central points by minimizing the sum of the quadratic distance between spins and their respective central points
\begin{equation} \label{eqn_kmeans}
	D(\vec{y}_1,\vec{y}_2,...,\vec{y}_k)=\sum_{i=1}^{m} \sum_{\vec{x}_j\in S_i}||\vec{x}_j-\vec{y}_i||^2.
\end{equation}
The positions $\vec{y}_i$ of the central points are varied to form $m$ partitions $S_i$ of proximate spins. During the optimization, spins can change the partitions, i.e., the respective central points that they are assigned to. Therefore, the size of individual partitions is not fixed.

To form partitions with the same number of spins, illustrated in Fig.~\ref{fig_partition}, we use a variant of the k-means algorithm, namely the \textit{constrained k-means algorithm}~\cite{Levy-Kramer_k-means-constrained_2018}. Using this algorithm, the optimization routine for the above quantity in Eq.~\eqref{eqn_kmeans} is subject to the additional constraints that forces the partitions to be of the same size.

\begin{figure} [t]
	\includegraphics[width=0.7\columnwidth]{figure_partitioning_2.png}
	\caption{Illustration of different partitioning of a system: a) sketch of the spin system: a central spin (red) interacting with a spin bath (blue). b) - d) the spin bath is partitioned into local subsystems (partitions) each containing one (b), two (c) and three (d) spins. The partitions are indicated by ovals and different colors. The configuration in (b) corresponds to the conventional CCE method ($K=1$), whereas (c) and (d) correspond to the novel pCCE method with $K=2$ and $K=3$, respectively.}
	\label{fig_partition}
\end{figure}

After finding the partitions, they represent indivisible units for the pCCE method. Clusters are formed based on the partitions. For example, pCCE(2,4) means that clusters are formed from one or two partitions with 4 and $2\cdot4=8$ spins, respectively.

When investigating P1-center baths, we choose the following partitioning of the bath to increase the efficiency of the method: Each of the five subsystems within the P1-center bath introduced above is partitioned separately. This procedure avoids the situation when spins from different subsystems (which do not flip-flop with each other) enter the same partition. We checked whether this partitioning of the bath yields accurate results by partitioning (for some of the settings) the whole P1-center bath without paying attention to the subgroups. Both approaches lead to similar results.

The pCCE method is compatible with the gCCE approach (see, for example, Refs.~\cite{onizhuk_probing_2021,maile}), which includes the central spin in each cluster. In fact, in order to use canonical typicality mentioned above, we used in our calculations the gCCE approach, which is equivalent to the conventional CCE method for the systems considered.

To obtain the quantum-mechanical time-evolution operator from a Hamiltonian, we use the \textit{scipy.linalg.expm} function from the \textit{scipy} library. This function is based on the method described in Ref.~\cite{scipy}.

\section{Structural comparison between pCCE and CCE}
In this section, we compare the pCCE and CCE methods regarding the clusters included in the calculation. We start with an explicit example, which is followed by a more general discussion.

\subsection{An explicit example}
Let us explicitly consider the clusters included in the numerical calculations in different orders of pCCE and CCE for the six-spin bath sketched in Fig.~\ref{fig_partition}. When using CCE$N$, clusters can be formed from individual spins, cf. Fig~\ref{fig_partition} (b). We compare these results to pCCE($N$,2), where the corresponding partitions are shown in Fig~\ref{fig_partition} (c). We use the notation $\widetilde{\cal L}_{(n_1,n_2,...)}$, where $n_1$, $n_2$, ... denote the spins included in the cluster according to the numbering provided in Fig~\ref{fig_partition} (b).

For CCE1, we obtain
\begin{equation} \label{eqn_cce1}
	\langle M_x(2\tau)\rangle_\text{CCE1}=\langle\langle\widetilde{\cal L}_{(1)}\widetilde{\cal L}_{(2)}\widetilde{\cal L}_{(3)}\widetilde{\cal L}_{(4)}\widetilde{\cal L}_{(5)}\widetilde{\cal L}_{(6)}\rangle\rangle,
\end{equation}
where the contribution of each individual spin is included $\widetilde{\cal L}_{(n_1)}=\langle M_x(2\tau)\rangle_{(n_1)}$. In CCE2, the contribution of all possible spin pairs is also taken into account
\begin{equation} \label{eqn_cce2}
	\begin{split}
		\langle M_x(2\tau)\rangle_\text{CCE2} = \langle\langle\widetilde{\cal L}_{(1)}\widetilde{\cal L}_{(2)}\widetilde{\cal L}_{(3)}\widetilde{\cal L}_{(4)}\widetilde{\cal L}_{(5)}\widetilde{\cal L}_{(6)}\widetilde{\cal L}_{(1,2)}\widetilde{\cal L}_{(1,3)}\widetilde{\cal L}_{(1,4)}\widetilde{\cal L}_{(1,5)}\widetilde{\cal L}_{(1,6)}\widetilde{\cal L}_{(2,3)}\widetilde{\cal L}_{(2,4)}\\\widetilde{\cal L}_{(2,5)}\widetilde{\cal L}_{(2,6)}\widetilde{\cal L}_{(3,4)}\widetilde{\cal L}_{(3,5)}\widetilde{\cal L}_{(3,6)}\widetilde{\cal L}_{(4,5)}\widetilde{\cal L}_{(4,6)}\widetilde{\cal L}_{(5,6)}\rangle\rangle,
	\end{split}    
\end{equation}
where the two-spin terms read $\widetilde{\cal L}_{(n_1,n_2)}=\frac{\langle M_x(2\tau)\rangle_{(n_1,n_2)}}{\langle M_x(2\tau)\rangle_{(n_1)}\langle M_x(2\tau)\rangle_{(n_2)}}$, i.e., the two-spin terms must be normalized by the single-spin contributions.

In pCCE(1,2), only three individual partitions of size $K=2$ are included
\begin{equation} \label{eqn_pcce(1,2)}
	\langle M_x(2\tau)\rangle_\text{pCCE(1,2)} =\langle\langle\widetilde{\cal L}'_{(1,2)}\widetilde{\cal L}'_{(3,4)}\widetilde{\cal L}'_{(5,6)} \rangle\rangle,
\end{equation}
where the individual terms do not need to be normalized because the partitions are the basic units for pCCE, i.e., $\widetilde{\cal L}'_{(n_1,n_2)}=\langle M_x(2\tau)\rangle_{(n_1,n_2)}$. This is why we can rewrite Eq.~\eqref{eqn_pcce(1,2)} as
\begin{equation} \label{eqn_pcce_cce_translation}
	\langle M_x(2\tau)\rangle_\text{pCCE(1,2)} = \langle\langle\widetilde{\cal L}_{(1)}\widetilde{\cal L}_{(2)}\widetilde{\cal L}_{(3)}\widetilde{\cal L}_{(4)}\widetilde{\cal L}_{(5)}\widetilde{\cal L}_{(6)}\widetilde{\cal L}_{(1,2)}\widetilde{\cal L}_{(3,4)}\widetilde{\cal L}_{(5,6)}\rangle\rangle,
\end{equation}
which means that pCCE(1,2) goes beyond CCE1 (cf. Eq.~\eqref{eqn_cce1}) by including the most important two-spin terms.

In pCCE(2,2), the corresponding expansion reads
\begin{equation} \label{eqn_pcce(2,2)}
	\langle M_x(2\tau)\rangle_\text{pCCE(2,2)} = \langle\langle\widetilde{\cal L}'_{(1,2)}\widetilde{\cal L}'_{(3,4)}\widetilde{\cal L}'_{(5,6)} \widetilde{\cal L}'_{(1,2,3,4)}\widetilde{\cal L}'_{(1,2,5,6)}\widetilde{\cal L}'_{(3,4,5,6)}\rangle\rangle,
\end{equation}
where 
\begin{equation}
	\widetilde{\cal L}'_{(n_1,n_2,n_3,n_4)}=\frac{\langle M_x(2\tau)\rangle_{(n_1,n_2,n_3,n_4)}}{\langle M_x(2\tau)\rangle_{(n_1,n_2)}\langle M_x(2\tau)\rangle_{(n_3,n_4)}}.
\end{equation}
In the denominator of the above expression, only the partitions forming the four-spin cluster $(n_1,n_2,n_3,n_4)$ enter, while, in the conventional CCE, a cluster of 4 spins needs to be normalized with all possible subclusters including clusters containing 1, 2 and 3 spins. An important implication of this modified normalization is that we obtain similar to Eq.~\eqref{eqn_pcce_cce_translation} now
\begin{equation} \label{eqn_pcce_cce_translation2}
	\begin{split}
		\langle M_x(2\tau)\rangle_\text{pCCE(2,2)} = \langle\langle\widetilde{\cal L}_\text{CCE2}\widetilde{\cal L}_{(1,2,3)}\widetilde{\cal L}_{(1,2,4)}\widetilde{\cal L}_{(1,3,4)}\widetilde{\cal L}_{(2,3,4)}\widetilde{\cal L}_{(3,4,5)}\widetilde{\cal L}_{(3,4,6)}\widetilde{\cal L}_{(3,5,6)}\widetilde{\cal L}_{(4,5,6)}\\\widetilde{\cal L}_{(1,2,5)}\widetilde{\cal L}_{(1,2,6)}\widetilde{\cal L}_{(1,5,6)}\widetilde{\cal L}_{(2,5,6)}\widetilde{\cal L}_{(1,2,3,4)}\widetilde{\cal L}_{(3,4,5,6)}\widetilde{\cal L}_{(1,2,5,6)}\rangle\rangle,
	\end{split}    
\end{equation}
where $\widetilde{\cal L}_\text{CCE2}$ corresponds to the terms in Eq.~\eqref{eqn_cce2} entering the expression on the right-hand side within the brackets. Thus, by calculating the contribution of a four-spin cluster, also the contribution of the three-spin clusters are included without their explicit calculation.

In general, we conjecture that pCCE($N$,$K$) contains all contributions that enter CCE$N$ and also further higher-order terms up to order $NK$. When comparing pCCE($N$,$K$) with CCE$NK$, both expansions contain terms of order $NK$. While pCCE($N$,$K$) includes only the most important contributions from clusters which are the union of partitions, CCE$NK$ includes contributions from all possible clusters. For example, when comparing pCCE(2,2) in Eq.~\eqref{eqn_pcce_cce_translation2} and CCE4, CCE4 would also include all four-spin and three-spin clusters, which are not included in Eq.~\eqref{eqn_pcce_cce_translation2}.

\subsection{Comparison of the number of clusters}
The total time $T_\text{tot}$ required to obtain the final results for CCE or pCCE can be estimated as follows
\begin{equation} \label{eqn_ttot}
	T_\text{tot}\approx N_\text{sys}\cdot N_\text{clusters}\cdot N_\text{average}\cdot T(\text{size of largest clusters}),
\end{equation}
where $N_\text{sys}$ is the number of random spin distributions considered, $N_\text{clusters}$ is the number of largest clusters of the expansion, $N_\text{average}$ is the number of samples for the mean-field average and $T(\text{size of largest clusters})$ is the time required to calculate the contribution for one of the largest clusters. Only the largest clusters are considered in the above estimate because the calculation of their contribution usually covers most of the overall calculation time. While the factor $N_\text{sys}$ is the same for both methods, pCCE offers advantages in the remaining three factors entering Eq.~\eqref{eqn_ttot} as we show in the following.

To calculate the number of clusters $N_\text{clusters}$, we make the assumption that the dipole radius $r_\text{d}$ (defined below) is infinite, i.e., any spins can be combined in a cluster irrespective of the distance between the spins. Then, the number of largest clusters of size $NK$ is
\begin{equation}
	\begin{split}
		N_\text{clusters}=&\binom{\frac{N_s}{K}}{N}\text{ for pCCE($N$,$K$)}\\
		N_\text{clusters}=&\binom{N_s}{NK}\text{ for CCE$NK$},
	\end{split}    
\end{equation}
where $N_s$ is the total number of bath spins. For the typical case, when $N_s\gg KN$, $\binom{N_s/K}{N}\ll\binom{N_s}{NK}$. For $N_s=100$, we provide the numbers of clusters of different sizes included in pCCE and CCE in Table~\ref{table_cce_pcce_comp}. For the pCCE method, we also indicate how many clusters are effectively included without calculating their contribution separately, as explained in the previous section. For example, using pCCE(2,4), in addition to the 300 clusters of size 8, a large number of smaller clusters is included, cf. last column of Table~\ref{table_cce_pcce_comp}.

\begin{table}[t]
	\setlength\tabcolsep{0pt}
	\begin{tabular*}{\linewidth}{@{\extracolsep{\fill}} ccccccc }
		Cluster size&CCE4&CCE8&pCCE(2,2)&pCCE(2,2)&pCCE(2,4)&pCCE(2,4)\\
		
		&calculated&calculated&calculated&included&calculated&included\\
		\hline
		1&100&100&-&100&-&100 \\
		2&4\,950&4\,950&50&4950&-&4950 \\
		3&161\,700&161\,700&-&4900&-&14500 \\
		4&3\,921\,225&3\,921\,225&1225&1225&25&20425 \\
		5&-&$7.53\cdot10^7$&-&-&-&16800 \\
		6&-&$1.19\cdot10^9$&-&-&-&8400 \\
		7&-&$1.60\cdot10^{10}$&-&-&-&2400 \\
		8&-&$1.86\cdot10^{11}$&-&-&300&300
	\end{tabular*}
	\caption{Number of clusters of different size (number of spins included) calculated in CCE and pCCE for a bath of $N_s=100$ spins. For CCE, the number of calculated and incorporated clusters is the same (see text). In pCCE(2,$K$), only clusters of size $K$ and $2K$ are calculated but many other clusters are effectively included.}
	\label{table_cce_pcce_comp}
\end{table}

More importantly, when comparing the number of four-spin clusters for CCE4 and the number of eight-spin clusters for pCCE(2,4), the overall calculation time is expected to be smaller for pCCE(2,4) than for CCE4 because the ratio of these numbers is approximately $10^4$, cf. Table~\ref{table_cce_pcce_comp}, which is much larger than the expected scaling of the computational time $4^n$ with the spin number $n$ ($4^4=256$). This discrepancy in the number of clusters becomes even stronger, when $N$ is further increased. Even after introducing a finite dipole radius $r_\text{d}$ (see below) and, thereby, reducing the number of four-spin clusters for CCE4, we anticipate that the pCCE calculation terminates faster. We expect pCCE(2,4) to yield results comparable to CCE4 (or better) in a shorter time. 


When considering $N_\text{average}$, pCCE also offers an advantage here, as shown above. For example, when comparing CCE4 and pCCE(2,4), we expect that CCE4 requires at least 100 times more samples for the mean-field average than pCCE(2,4) and this factor further grows when increasing $N$.

Finally, for the last term in Eq.~\eqref{eqn_ttot}, pCCE also offers an advantage because the cluster sizes are usually larger for pCCE($N,K$) than for CCE$N$. This allows, in turn, to use canonical typicality~\cite{gemmer,canonical_typicality,popescu} for the bath states of large clusters. Quantum typicality implies that randomly chosen quantum states typically lead to good approximations of infinite-temperature expectation values for the bath spins. The calculations need to be averaged over sufficiently many randomly chosen initial quantum states for the P1-center bath to ensure convergence. This allows to use the wave-function formalism instead of the density matrices, thereby requiring less computational resources.

\section{Comparison of results of the {pCCE} method and exact simulations for small systems}
To study the convergence behavior of the pCCE method, we calculate the spin coherence $\langle M_x(2\tau)\rangle$ of NV spins in small baths of 20 electron spins-1/2. Here, we discuss results for three different spatial spin distributions, a two-dimensional spin lattice and two random three-dimensional baths, and compare the numerical pCCE results with results of an exact simulation. The exact simulations were performed using the Suzuki-Trotter method~\cite{suzuki}.

As the first example, we discuss results for a spin-lattice system with bath spins being placed on a two-dimensional lattice with nearest-neighbor distance of 20\,nm. The results obtained with the pCCE method for different partition size $K$ are shown in Fig.~\ref{fig_test} (a). Increasing the partition size $K$, the results converge towards the exact solution; an exception is pCCE(2,2) showing unphysical behavior. This behavior is suppressed when using internal mean-field averaging, as illustrated in Fig.~\ref{fig_test} (a) by the green curve.

\begin{figure*}[b]
	\includegraphics[width=0.99\linewidth, height=5cm]{figure_example_graphs2.png}
	\caption{Spin coherence $\langle M_x(2\tau)\rangle$ decay of the NV electron spin in three different small baths of 20 electron spins-1/2: (a) a two-dimensional spin lattice, and (b) - (c) three-dimensional spin baths with random spin distributions at concentration 1\,ppm (see text). In all three cases, the pCCE results converge towards the exact results when increasing $K$. To demonstrate the effect of internal mean-field averaging, we also include results with and without internal mean-field average. Results obtained with internal averaging are labeled as ``int. avg.''. In (b), already the CCE2 (equivalent to pCCE(2,1)) results show good agreement with exact results, whereas, in (c), there are sizable deviations between CCE2 and the exact solution.}
	\label{fig_test}
\end{figure*}

The deviation of CCE2 results from the exact solution depends on the particular spin system considered~\cite{Witzel2012}. Sparse spin systems can cause vastly different decays depending on the spatial spin arrangement. To illustrate this, we chose two different three-dimensional systems and show results for these systems in Fig.~\ref{fig_test} part (b) and (c). Both spin systems were obtained by placing the bath electron spins on a diamond lattice at a concentration of 1 ppm and then selecting the 20 closest spins to the central NV center. For the first system in part (b), the CCE2 results deviate only slightly from the exact solution. When increasing the partition size $K$, these small deviations are reduced. In the second system in part (c), the deviations of CCE2 (equivalent to pCCE(2,1)) are larger. Here, we also observe a convergence to the exact results when increasing $K$. For all systems considered, pCCE(2,4) shows good agreement with the exact solution on the relevant timescales.

For the exact simulations and also for the pCCE method with $K\approx4$, we use canonical typicality for the P1-center bath state, which means that randomly chosen quantum states typically lead to good approximations of infinite-temperature expectation values~\cite{gemmer,canonical_typicality,popescu} for the bath spins. The calculations are averaged over sufficiently many randomly chosen initial quantum states for the P1-center bath to ensure convergence.

\section{Parameters of the pCCE method for large spin baths}
In this section, we discuss the values of parameters for the pCCE method used in our calculations.

\subsection{Bath parameters}
In our numerical simulations, we first create a very large diamond lattice with the NV center in the center of this lattice and distribute P1 centers randomly on this lattice. P1-center spins sufficiently far away from the NV spin should have a negligible effect on the spin-coherence decay of the NV spin. Therefore, we limit the number of P1-center spins included in the simulation. We do this by introducing a bath radius $r_\text{b}$ which defines a sphere around the NV spin. Only spins within this sphere are considered dynamical.

Another parameter is the dipole radius $r_\text{d}$ which limits the distance between the partitions to form a joint cluster. The idea behind the dipole radius $r_\text{d}$ is that flip-flop transitions between spins from partitions far away from each other can be neglected. For the CCE approach, this corresponds to the distance between the respective spins. In the pCCE approach, the distance between two partitions is taken as the distance between the respective center points obtained from the constrained k-means algorithm (see above). Both parameters $r_\text{b}$ and $r_\text{d}$ need to be adjusted to each system considered.

Around the sphere of radius $r_\text{b}$ described above, we include further P1-center spins within a spherical shell of thickness $\frac{2}{3} r_\text{d}$ on the mean-field level, i.e., they are static. These additional spins are particularly important for spins close to the boundary of the sphere of radius of $r_\text{b}$. This means that, in total, we cut out a sphere of radius $r_\text{b} + \frac{2}{3} r_\text{d}$ from the large diamond lattice mentioned above and consider only spins within this sphere in our simulations.

\begin{figure} [b]
	\includegraphics[width=0.6\linewidth]{convergence_r_graphs.png} 
	\caption{Evidence for the convergence of parameters for the pCCE(2,4) method applied to a large bath of P1-center electron spins in a quasi-two-dimensional configuration at a concentration of 1\;ppm: a) the bath-radius $r_\text{b}$ and b) the dipole radius $r_\text{d}$. The values of $r_\text{b}$ and $r_\text{d}$ depend both on the layer thickness $L$ and the bath-spin concentration $\rho_\text{P1}$. In our calculations, we choose the values for the parameters $r_\text{b}$ and $r_\text{d}$ corresponding to converged spin-coherence decay results, i.e., the spin-coherence decay does not change appreciably when increasing any of the parameters. For further details, see text.}
	\label{fig_test_radii}
\end{figure}

The values of the above parameters $r_{\text{b}}$ and $r_\text{d}$ were determined in convergence studies. An example is illustrated in Fig.~\ref{fig_test_radii}, where the spin-coherence decay $\langle M_x(2\tau)\rangle$ of the NV spin is shown for a P1-center spin bath at concentration 1\,ppm using pCCE(2,4). The spin-coherence decay $\langle M_x(2\tau)\rangle$ is averaged over 100 randomly distributed spin systems and 20 mean-field spin configurations in internal averaging. The bath radius is increased in steps of 10\,nm starting from 60\,nm, see Fig.~\ref{fig_test_radii} a. The dipole radius is increased in steps of 10\% starting from the initial value of 64.9\,nm, see Fig.~\ref{fig_test_radii} b. In both calculations, the other parameter was kept constant at its initial value. The resulting graphs do not show any significant changes when increasing the parameter values. We thus use the values $r_{\text{b}}=60$\,nm and $r_{\text{d}}=64.9$\,nm in our simulations for this particular system. Within the radius $r_{\text{b}}=60$\,nm, there are 140 bath spins, which corresponds to the number of P1 centers. Small fluctuations of results shown in Fig.~\ref{fig_test_radii} arise presumably due to a finite number of samples for the mean-field average. The chosen parameter values are similar to those used in other studies of NV spins in P1-center baths for CCE2, see for example Ref.~\cite{park_decoherence_2022}. 

When considering other concentrations $\rho_{\text{P1}}$, the bath radius $r_{\text{b}}$ is chosen such that there are at least 140 spins, i.e., P1 centers, within the radius. To assure partitions of the same size within each P1-bath subgroup (discussed above), further spins are added to each subgroup until the total number of spins is divisible by the partition size $K$.

The dipole radius $r_{\text{d}}$ is, in general, a function of the concentration $\rho_{\text{P1}}$. We thus define
\begin{equation} \label{eqn_rdip1}
	\tilde{r}_\text{d}(\rho_{\text{P1}})\equiv r_\text{d,1}\left(\frac{\rho_{\text{P1}}}{1\ \text{ppm}}\right)^{-1/3},
\end{equation}
where $r_{\text{d,1}}$ is the dipole radius at 1\,ppm. To incorporate the effect of different spatial dimensionalities of the bath, we set
\begin{equation} \label{eqn_rdip2}
	r_\text{d}(L,\rho_{\text{P1}})=\tilde{r}_\text{d}(\rho_{\text{P1}}) \max \left[ 1,\sqrt{\frac{2\tilde{r}_\text{d}(\rho_{\text{P1}})}{L}} \right],
\end{equation}
where the last term on right-hand-side makes $r_\text{d}$ larger for quasi-two-dimensional layers. We also use the same values of $r_\text{b}$ and $r_\text{d}$ when leaving out the hyperfine interaction in the P1 centers.

\begin{figure}[b]
	\includegraphics[width=0.8\linewidth]{figure_con_order.png}
	\caption{Convergence of the pCCE method with respect to the partition size $K$ and the number of samples for the internal mean-field average for different spin baths. a) An electron spin-1/2 bath in a quasi-two-dimensional layer of thickness $L=5$\,nm at a concentration of 1\,ppm. At larger $K$, results for different numbers of samples for the internal average are also shown to demonstrate the convergence with the number of samples. b) A P1-center spin bath in a quasi-two-dimensional layer ($L=30$\,nm) at a concentration of $\rho_\text{P1}=1$\,ppm. Fluctuation of results of pCCE(2,3) likely arise due to insufficient internal mean-field averaging. In a and b, the results for $K=2$ and $K=3$ were averaged internally over 100 and 50 mean-field samples, respectively.}
	\label{fig_con_order}
\end{figure}

\subsection{Partition size $K$ and the number of samples for the mean-field average}
We first consider a quasi-two-dimensional system of electron spins-1/2 with layer thickness $L=5$\,nm and concentration of 1\,ppm and calculate the spin-coherence decay $\langle M_x(2\tau)\rangle$. Figure~\ref{fig_con_order}\,a shows results for different partition sizes $K$ and different number of samples for the internal mean-field average. The results indicate that the convergence is achieved for partition size $K=4$ and 20 samples for the internal mean-field average.

The decay of $\langle M_x(2\tau)\rangle$ becomes faster for larger $K$. This implies that, when using CCE2, the resulting $T_2$ time is too large in the quasi-two-dimensional case. On the other hand, when considering three-dimensional baths of electron spins-1/2, the pCCE(2,$K$) results for all $K\geq1$ demonstrate similar behavior (not shown here) which indicates that already pCCE(2,1) (equivalent to CCE2) provides accurate results. For the sake of consistency, we use pCCE(2,4) also for these systems with the same number of samples for the internal average.

For the P1-center spin baths, no significant difference between pCCE(2,1) and pCCE(2,4) was observed for all layers, as indicated in Fig.~\ref{fig_con_order} b. We attribute this to the strong suppression of flip-flop transitions in the bath due to the hyperfine interaction, see the first section of this supplemental material. The CCE2 method thus yields sufficiently accurate results for P1-center spin baths. However, this is strictly speaking valid only for the spin-coherence decay averaged over many different random spin distributions. This does not imply that, for each individual spin system, CCE2 yields sufficiently accurate results, cf. Ref.~\cite{Witzel2012}. Using pCCE, accurate results can also be obtained for individual systems. It is also noteworthy, that, for establishing the validity of CCE2 approach for P1-center spin baths, the extensive use of the pCCE method was necessary.

\section{Fitting the stretched-exponential function}
\begin{figure}[b]
	\includegraphics[width=0.8\linewidth]{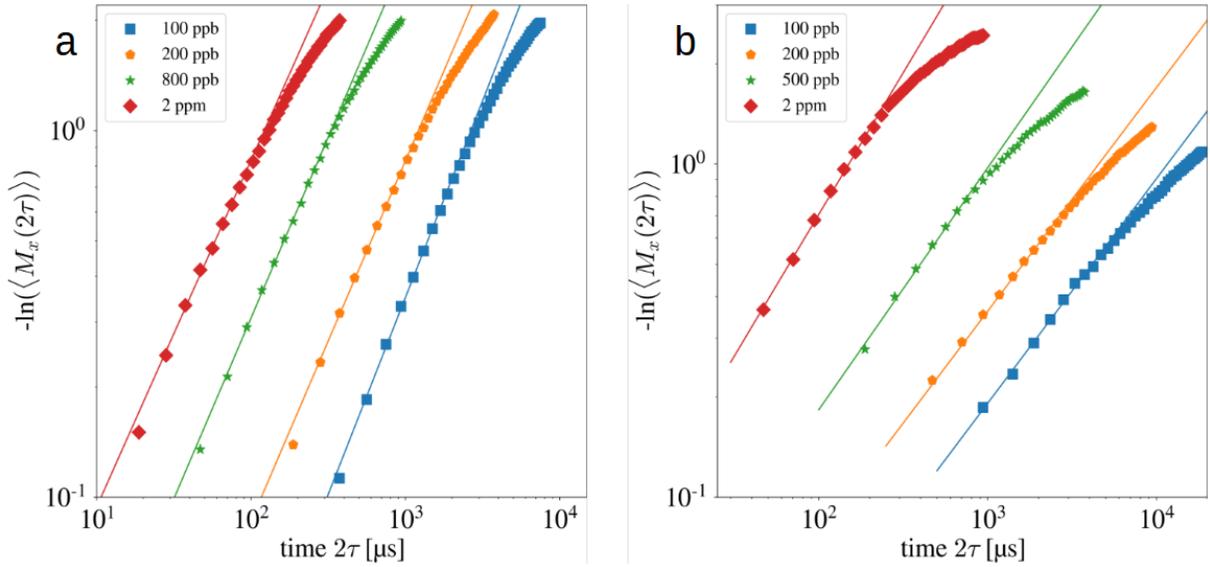}
	\caption{Illustration of the fitting of the stretched-exponential function in Eq.~(2) to the numerical results shown in Fig.~3 at $L=240$\;nm (a) and $L=60$\;nm (b). The fitting parameters are the stretched-exponential parameter $p$ and the $T_2$ time. The slope of the linear curves and, hence, the parameter $p$ increases with increasing P1-center spin concentration $\rho_\text{P1}$ as indicated in the legend. The time window for the fit was adjusted for each individual curve separately.}
	\label{fig_fit}
\end{figure}

The stretched-exponential parameter $p$ and the characteristic $T_2$ time are extracted by fitting the stretched-exponential function in Eq.~(2) to the short-time decay of the spin coherence $\langle M_x(2\tau)\rangle$. We perform a linear fit to $\ln(-\ln(\langle M_x(2\tau)\rangle))$ as a function of $\ln(t)$, from which we extract the slope corresponding to $p$, as illustrated in Fig.~\ref{fig_fit} for P1-center spin baths. Additionally, we extract the y-axis offset $d$. The $T_2$ time can then be calculated via $T_2 = \exp\left(-\frac{d}{p}\right)$. We adjust the time window for the linear fit for each decay curve separately. We extend the time window to late times until a significant bending towards a smaller slope is observed. This bending usually occurs when $\langle M_x(2\tau)\rangle$ decayed below 0.6.

\begin{figure}[t]
	\includegraphics[width=0.8\linewidth]{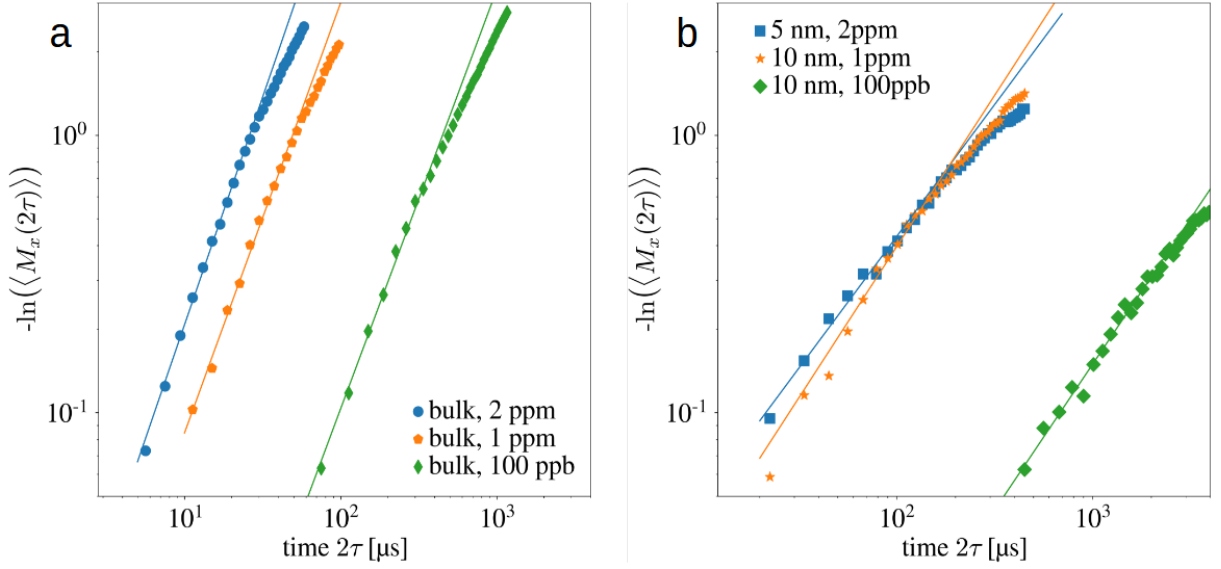}
	\caption{Illustration of the fitting of the stretched-exponential function in Eq.~(2) to the numerical results obtained for P1 centers without the hyperfine interaction for three-dimensional (a) and quasi-two-dimensional (b) diamond layers, cf. Table~1 of the main article. The layer thickness $L$ and the P1-center concentration $\rho_\text{P1}$ are indicated in the legends. The fitting parameters are the stretched-exponential parameter $p$ and the $T_2$ time. The time window for the fit was adjusted for each individual curve separately.}
	\label{fig_fit_e}
\end{figure}

Similar behavior is observed for the baths of P1 centers without the hyperfine interaction, i.e., baths of electron spins-1/2. The resulting curves and corresponding fits are shown in Fig.~\ref{fig_fit_e}. The transitions from three-dimensional to quasi-two-dimensional systems occur at lower layer thickness compared to the P1-center spin bath. The reason for this is the absence of the suppression of the flip-flop transitions by the hyperfine interaction and, hence, different spin dynamics in the bath. Therefore, in Table~1 of the main text, we considered for the quasi-two-dimensional configuration for P1 centers a layer thickness of $L=30\,$nm at 2\,ppm and 1\,ppm and $L=60\,$nm at 0.1\,ppm, and, for P1 centers without the hyperfine interaction, $L=5\,$nm at 2\,ppm and $L=10\,$nm at 0.1\,ppm and 1\,ppm.

In general, we find that the decay of $\langle M_x(2\tau)\rangle$ comprises three regimes. At very small times, we observe a slope corresponding to a higher stretched-exponential parameter $p>1.5$. This is followed by the regime, which we refer to as the short-time decay. Here, we perform the linear fit as indicated in Figs.~\ref{fig_fit} and~\ref{fig_fit_e}. Finally, the late-time decay is characterized by a lower stretched exponential, which can also be recognized by the bending of the curves in Fig.~3 of the main text, cf. Ref.~\cite{davis_probing_2023}.

The above three regimes complicate the choice for a suitable region for the fit of the short-time decay of $\langle M_x(2\tau)\rangle$. Thus, it is difficult to reliably identify the beginning and the end of this region such that the obtained slope of the curves corresponding to the stretched-exponential parameters $p$ together with the characteristic $T_2$ time must be considered an approximation. 

\section{Estimation of errors}
The P1-center spin baths considered here are sparse, which means that the P1-center concentration $\rho_\text{P1}$ is much smaller than the carbon-atom concentration in pure diamond. This leads to large variations of the spatial P1-center spin distribution and, hence, to large variations in the spin-coherence decay. In Figure~\ref{fig_spaghetti}, we show all 100 individual spin-coherence decays together with the mean value (black dots) for a quasi two- (b) and a three-dimensional (a) spin bath. The error bars indicate the standard deviation at the individual times. The standard-deviation of the mean is by the factor $\sqrt{100}=10$ smaller and, hence, comparable or smaller than the point size used in the figure.

\begin{figure}[h]
	\includegraphics[width=0.8\linewidth]{spaghetti_strands_error_4.png}
	\caption{Spin-coherence decay for each individual spatial bath-spin distribution is shown for (a) a three-dimensional and (b) a quasi-two-dimensional P1-center spin bath. The mean is indicated by the black dots and the error bars indicate the standard deviation resulting from the large variation of the decay curves.}
	\label{fig_spaghetti}
\end{figure}

To estimate the errors of the $T_2$ time and the stretched-exponential parameter $p$, we consider the covariance matrix obtained from the linear fit to the spin-coherence decay in a double-logarithmic plot as detailed in the previous section. From this covariance matrix, the standard deviation of the fitting parameters is calculated. In this fit, the standard deviation of the mean values of each data point mentioned above is included. Further, we add to the error 5~\% of the fitted values to account for the error arising from the choice of the time window for the fitting. The resulting error bars are shown in Fig.~4 of the main article.	
\end{document}